\input amstex
\documentstyle{amsppt}
\magnification=\magstephalf
%%%%%%%%%%%% changes to amsppt.sty %%%%%%%%%%%%%%%%%%%%%%%
 \addto\tenpoint{\baselineskip 15pt
  \abovedisplayskip18pt plus4.5pt minus9pt
  \belowdisplayskip\abovedisplayskip
  \abovedisplayshortskip0pt plus4.5pt
  \belowdisplayshortskip10.5pt plus4.5pt minus6pt}\tenpoint
\pagewidth{6.5truein} \pageheight{8.9truein}
\subheadskip\bigskipamount
\belowheadskip\bigskipamount
\aboveheadskip=3\bigskipamount
\catcode`\@=11
\def\output@{\shipout\vbox{%
 \ifrunheads@ \makeheadline \pagebody
       \else \pagebody \fi \makefootline 
 }%
 \advancepageno \ifnum\outputpenalty>-\@MM\else\dosupereject\fi}
\outer\def\subhead#1\endsubhead{\par\penaltyandskip@{-100}\subheadskip
  \noindent{\subheadfont@\ignorespaces#1\unskip\endgraf}\removelastskip
  \nobreak\medskip\noindent}
\outer\def\enddocument{\par% \par will do a runaway check for \endref
  \add@missing\endRefs
  \add@missing\endroster \add@missing\endproclaim
  \add@missing\enddefinition
  \add@missing\enddemo \add@missing\endremark \add@missing\endexample
 \ifmonograph@ % do nothing
 \else
 \vfill
 \nobreak
 \thetranslator@
 \count@\z@ \loop\ifnum\count@<\addresscount@\advance\count@\@ne
 \csname address\number\count@\endcsname
 \csname email\number\count@\endcsname
 \repeat
\fi
 \supereject\end}
\catcode`\@=\active
%%%%%%%%%%%%%%% other macros %%%%%%%%%%%%%%%%%%%%%%%%%%%%
\CenteredTagsOnSplits
\NoBlackBoxes
\nologo
\def\today{\ifcase\month\or
 January\or February\or March\or April\or May\or June\or
 July\or August\or September\or October\or November\or December\fi
 \space\number\day, \number\year}
\define\({\left(}
\define\){\right)}

\define\CC{{\Bbb C}}
\define\CP{{\Bbb C\Bbb P}}

\define\End{\operatorname{End}}

\define\HH{{\Bbb H}}
\define\Hom{\operatorname{Hom}}

\define\RR{{\Bbb R}}

\define\ZZ{{\Bbb Z}}
\define\[{\left[}
\define\]{\right]}

\define\chiup{\raise.5ex\hbox{$\chi$}}

\define\dbar{{\bar\partial}}

%\define\exertag #1#2{\removelastskip\bigskip\medskip\eightpoint\noindent%
%\hbox{\rm\ignorespaces#2\unskip} #1.\ }  
\define\exertag #1#2{#2\ #1}

\define\inv{^{-1}}
\define\mstrut{^{\vphantom{1*\prime y}}}
\define\protag#1 #2{#2\ #1}

\define\res#1{\negmedspace\bigm|_{#1}}
\define\temsquare{\raise3.5pt\hbox{\boxed{ }}}

\define\theprotag#1 #2{#2~#1}

\define\xca#1{\removelastskip\medskip\noindent{\smc%
#1\unskip.}\enspace\ignorespaces }

\define\zmod#1{\ZZ/#1\ZZ}

\redefine\Im{\operatorname{Im}}
\redefine\Re{\operatorname{Re}}

\NoRunningHeads % USE IN FINAL VERSION; THEN COMMENT OUT NEXT LINE

\define\AR{A_{\RR}}
\define\Abar{\overline{A}}
\define\Atil{\tilde{A}}
\define\Cx{$\CC^\times$}
\define\Kah{K\"ahler}
\define\Ktil{\tilde{K}}
\define\Ltil{\tilde{\Lambda }}
\define\Mtil{\tilde{M}}
\define\PP{\Bbb{P}}
\define\Qtil{\tilde{Q}}

\define\Sym{\operatorname{Sym}}
\define\TT{\Bbb T}
\define\Vbar{\overline{V}}
\define\cotM{T^*M}
\define\etah{\hat{\eta}}
\define\hyp{hyperk\"ahler}
\define\id{\operatorname{id}}
\define\mopl{\frac{-1}{\pi}\log}
\define\nabtil{\widetilde{\nabla}}
\define\omtil{\tilde{\omega }}
\define\proj{\pi^{(1,0)}}
\define\rhoh{\hat{\rho}}
\define\sbar{\bar{s}}
\define\scrA{\Cal{A}}
\define\scrF{\frak{F}}
\define\tpi{2\pi i}
\define\vect#1{\frac{\partial }{\partial #1}}

\refstyle{A}
\widestnumber\key{SSSSSSSSS}   % for widest bibliography name
\document

 \pretitle{$$\boxed{\boxed{\text{REVISED VERSION}}}$$\par\vskip 3pc}

	\topmatter
 \title\nofrills Special K\"ahler Manifolds \endtitle
 \author Daniel S. Freed  \endauthor
 \thanks The author is on leave from the Department of Mathematics at the
University of Texas at Austin, where he receives support from NSF grant
DMS-962698.  At the Institute for Advanced Study the author receives support
from NSF grants DMS-9304580 and DMS-9627351, the Harmon Duncombe Foundation,
and from the J. Seward Johnson Sr. Charitable Trust.\endthanks
 \affil Schools of Mathematics and Natural Sciences\\Institute for Advanced
Study\endaffil 
 \address {\it Permanent address:\/} Department of Mathematics, University of
Texas, Austin, TX 78712\endaddress
 \curraddr School of Mathematics, Institute for Advanced Study, Olden Lane,
Princeton, NJ 08540\endcurraddr
 \email dafr\@math.utexas.edu \endemail
 \date August 24, 1998\enddate
 %\dedicatory \enddedicatory
 \abstract We give an intrinsic definition of the special geometry which
arises in global $N=2$ supersymmetry in four dimensions.  The base of an
algebraic integrable system exhibits this geometry, and with an integrality
hypothesis any special \Kah\ manifold is so related to an integrable system.
The cotangent bundle of a special \Kah\ manifold carries a hyperk\"ahler
metric.  We also define special geometry in supergravity in terms of the
special geometry in global supersymmetry.\endabstract
	\endtopmatter

\document

Constraints on Riemannian metrics occur in many places in supersymmetry.  For
example, the requirement of extended supersymmetry in a two dimensional
$\sigma $-model constrains the target manifold to be \Kah\ or \hyp\ depending
on the amount of supersymmetry.  The scalars in supergravity theories are
often constrained to live on a particular homogeneous Riemannian manifold.
These sorts of special metrics---metrics with restricted holonomy group (such
as \Kah\ and \hyp\ metrics) and homogeneous metrics---are much studied by
Riemannian geometers, but there are situations in which we meet something
new.  One important example occurs in four dimensional gauge theories with
$N=2$~supersymmetry: the scalars in the vector multiplet lie in a {\it special
\Kah\ manifold\/}.  This is the case pertaining to global supersymmetry; when
coupled to $N=2$~supergravity in four dimensions the scalars lie in a {\it
projective special \Kah\ manifold\/}.\footnote{Physicists use the term
`special \Kah\ manifold' for both cases, and use words like `rigid' and
`local' to distinguish them.  Since these words have other connotations in
geometry, we adopt a different terminology.}  Notice that $N=1$~supersymmetry
already constrains the scalars to lie in a \Kah\ manifold, which must be
Hodge in the supergravity case.  Special geometry is the additional
constraint imposed by $N=2$~supersymmetry.
 
Special geometry appeared in the physics literature in~1984 in both global
supersymmetry~\cite{ST}, ~\cite{G} and supergravity~\cite{WP}.
Strominger~\cite{St} gave a coordinate-free definition in the supergravity
case.  Projective special \Kah\ manifolds are important in mirror symmetry,
as explained by Candelas and de la Ossa~\cite{CO}.  Special \Kah\ manifolds
in global supersymmetry have received more attention recently due to their
prominent role in the seminal work of Seiberg and Witten on $N=2$
supersymmetric Yang-Mills theories~\cite{SW1}, ~\cite{SW2}.  See~\cite{F}
and~\cite{CRTP} for recent discussions of special geometry and for extensive
references.
 
In this paper we introduce an {\it intrinsic\/}\footnote{{\it Intrinsic\/}
geometry concerns the tangent bundle and associated bundles, whereas {\it
extrinsic\/} geometry involves bundles not constructed directly from the
coordinate charts of a manifold.  Definitions of special geometry in the
physics literature are not intrinsic in this sense.} definition of special
geometry: A special \Kah\ structure is a flat connection on the {\it tangent
bundle\/} of a \Kah\ manifold.  The crucial condition is expressed
in~\thetag{1.2}.  From it follow the usual equations for special coordinates,
the holomorphic prepotential, the \Kah\ potential, etc.  We recount this
in~\S{1}, where we also define this geometry in terms of a holomorphic cubic
form.  In~\S{2} we construct a \hyp\ metric on the cotangent bundle of a
special \Kah\ manifold.  A local version of this result appears in the
physics literature~\cite{CFG}.  It seems likely that there is actually a one
parameter family of \hyp\ metrics of which the one we construct is a limiting
case (see~\cite{SW3}), but we have not pursued that here.  In~\S{3} we prove
the assertion made by Donagi and Witten~\cite{DW} that with a suitable
integrality hypothesis a special \Kah\ manifold parametrizes an algebraic
completely integrable system.  As a consequence the total space of an
algebraic integrable system carries a \hyp\ metric.  The usual definition of
a projective special \Kah\ manifold is based on a particular type of
variation of Hodge structure, which was first studied by Bryant and
Griffiths~\cite{BG}.  Our main observation here is that a projective special
\Kah\ structure on a Hodge manifold~$M$ of dimension~$n$ induces a special
{\it pseudo\Kah\/} structure of Lorentz type on a closely related
manifold~$\Mtil$ of dimension~$n+1$.  ($\Mtil$~is the total space of the
Hodge line bundle with the zero section omitted.)  With a suitable
integrality hypothesis the associated intermediate Jacobians are an
integrable system and carry a \hyp\ metric, results obtained
previously~\cite{DM2}, ~\cite{C}.  Finally, in~\S{5} we make some brief
comments on the physics (in the case of global supersymmetry).  We explain
that supersymmetry combined with the quantization of electric and magnetic
charges leads to the conclusion that integrable systems must enter into the
low energy description of $N=2$~supersymmetric gauge theories.

As mentioned above, the base of an algebraic integrable system is a special
\Kah\ manifold.  This is, I believe, the proper context for special \Kah\
geometry.  There are many examples of algebraic integrable systems, and
hopefully this excuses the paucity of examples presented here.
 
As mentioned in the footnote on the previous page, our terminology differs
from that in the physics literature.  We include the following table to aid
in translation: 
 
 \bigskip
\define\bigstrut{\vrule height12pt depth3pt width0pt} 
 \centerline{\eightpoint
  $$ \vtop{\offinterlineskip
      \halign{\hfil#\hfil\bigstrut\quad&\vrule\quad \hfil #\hfil\cr
      Our Terminology&Physics Literature\cr
      \omit&\vrule height2pt width0pt\cr
      \noalign{\hrule}
      \omit&\vrule height1pt width0pt\cr
      \noalign{\hrule}
      \omit&\vrule height3pt width0pt\cr
      Special \Kah & Rigid Special \Kah\cr 
      \omit&\vrule height2pt width0pt\cr
      &(vector multiplets in global $N=2$ supersymmetry)\cr
      \omit&\vrule height3pt width0pt\cr
      \noalign{\hrule}
      \omit&\vrule height3pt width0pt\cr
      Projective Special \Kah & (Local) Special \Kah\cr
      &(vector multiplets in $N=2$ supergravity)\cr
      }} $$
}
\bigskip

This paper grew out of a seminar talk explaining~\cite{DW}, and it had a long
gestation period since.  During that time I benefited from conversations and
lectures by many colleagues, including Jacques Distler, Ron Donagi, Nigel
Hitchin, Graeme Segal, Nathan Seiberg, Karen Uhlenbeck, and Edward Witten.
From the first version of the paper I received helpful remarks from Vicente
Cort\'es, James Gates, Zhiqin Lu, Simon Salamon, and the referees.  I thank
them all.

 \newpage
 \head
 \S{1} Definition and Basic Properties
 \endhead
 \comment
 lasteqno 1@ 40
 \endcomment

We introduce the following definition.

        \definition{\protag{1.1} {Definition}}
 Let $M$ be a \Kah\ manifold with \Kah\ form~$\omega $.  A {\it special
\Kah\/} structure on~$M$ is a real flat torsionfree symplectic
connection~$\nabla $ satisfying
       $$ d_\nabla I=0, \tag{1.2} $$
where $I$~is the complex structure on~$M$.
        \enddefinition

First we examine the consequences of the connection on the underlying real
symplectic structure on~$M$.  The connection~$\nabla $ determines an
extension of the de Rham complex 
  $$ 0 @>>> \Omega ^0(TM) @>d_\nabla=\nabla >> \Omega ^1(TM) @>d_\nabla >> 
     \Omega ^2(TM) @>d_\nabla >> \cdots \tag{1.3} $$
The flatness is the condition $d_\nabla ^2=0$.  Note that the Poincar\'e
lemma holds for~\thetag{1.3}: a closed $TM$-valued form is locally exact.
The torsionfree condition may be expressed by 
  $$  d_\nabla (\id)=0, \tag{1.4} $$
where $\id\in \Omega ^1(TM)$ is the identity endomorphism of~$TM$.  Now if
$\{\xi _\alpha \}$ is a flat local framing of~$M$ with dual
coframing~$\{\theta ^\alpha \}$, then \thetag{1.4}~implies $d\theta ^\alpha
=0$, whence $\theta ^\alpha =dt^\alpha $ for some local coordinate
functions~$t^\alpha $.\footnote{For simplicity we always choose our
coordinate systems to be defined on {\it connected\/} open sets, and we allow
the domains of the coordinate systems to shrink when necessary.}  Since
$\nabla \omega =0$ we can choose these coordinates to be {\it Darboux\/};
that is, the coordinate functions are~$x^i,y_j\;(i,j=1,\dots ,n=\dim _\CC M)$
with
  $$ \omega =dx^i\wedge dy_i. \tag{1.5} $$
 
Summarizing, a flat torsionfree symplectic connection~$\nabla $ is equivalent
to a flat symplectic structure on~$M$.  This is a covering by {\it flat
Darboux coordinate systems\/}~$\{x^i,y_j\}$ whose transition functions are of
the form
  $$ \pmatrix x\\y \endpmatrix = P\pmatrix \tilde{x}\\\tilde{y} \endpmatrix +
     \pmatrix a\\b \endpmatrix,\qquad P\in Sp(2n;\RR),\quad a,b\in
     \RR^n. \tag{1.6} $$
(The coordinates are ``flat'' since $\nabla dx^i=\nabla dy_j=0$.)
Equation~\thetag{1.5} is valid in any flat Darboux coordinate system. 
 
The compatibility with the complex structure is expressed\footnote{We give a
characterization in terms of coordinates in \theprotag{1.25} {Proposition}
below.}  by~\thetag{1.2}, or equivalently by
  $$ d_\nabla \pi ^{(1,0)} = 0, \tag{1.7} $$
where $\proj\in \Omega ^{1,0}(T_\CC M)$ is projection onto the $(1,0)$~part
of the complexified tangent bundle.  The Poincar\'e lemma ensures that
locally we can find a complex vector field~$\zeta $ with 
  $$ \nabla \zeta =\proj. \tag{1.8} $$
Note that $\zeta $~is unique up to a flat complex vector field.  Also,
$\zeta$~is not necessarily holomorphic.  Let $\{x^i,y_j\}$ be a flat Darboux
coordinate system and write
  $$ \zeta =\frac 12 \bigl(z^i\vect{x^i} - w_j\vect{y_j} \bigr) \tag{1.9} $$
for some complex functions~$z^i,w_j$.  (The choice of sign and the
factor~`1/2' yield standard formulas for~$M=\CC^n$.)  Since $\proj$~has
type~$(1,0)$, equation~\thetag{1.9} implies that $z^i,w_j$~are holomorphic
functions and
  $$ \proj = \frac 12\bigl(dz^i\otimes \vect{x^i} - dw_j\otimes \vect{y_j}
     \bigr). \tag{1.10} $$ %KEEPTAG
It follows that 
  $$ \Re(dz^i) = dx^i,\qquad \Re(dw_j) = -dy_j. \tag{1.11} $$ %KEEPTAG
In particular, $\{z^i\}$~is a local holomorphic coordinate system
on~$M$.\footnote{Of course, so is~$\{w_j\}$.  We call $\{z^i\}$ and $\{w_j\}$
{\it conjugate\/} coordinate systems (\theprotag{1.37} {Definition}).}  We
easily compute
  $$ \vect{z^i} = \frac 12\bigl(\vect{x^i}- \tau _{ij}\vect{y_j} \bigr),
     \tag{1.12} $$ %KEEPTAG
where 
  $$ \tau _{ij} = \frac{\partial w_j}{\partial z^i}. \tag{1.13} $$ %KEEPTAG
Now the fact that $\omega $~has type~$(1,1)$ implies that~$\tau _{ij}=\tau
_{ji}$, and so there is a (local) holomorphic function~$\scrF$, determined
up to a constant, so that 
  $$ w_j=\frac{\partial \scrF}{\partial z^j},\qquad \tau _{ij} =
     \frac{\partial ^2\scrF}{\partial z^i\partial z^j}. \tag{1.14} $$ %KEEPTAG
$\scrF$~is called the {\it holomorphic prepotential\/}.  It determines a
\Kah\ potential 
  $$ K = \frac 12\Im\bigl(\frac{\partial \scrF}{\partial z^i}\bar{z}^i\bigr)
     = \frac 12\Im(w_i\bar{z}^i), \tag{1.15} $$ %KEEPTAG
and in terms of this data the \Kah\ form is
  $$ \omega =\sqrt{-1}\partial \bar{\partial}K = \frac
     {\sqrt{-1}}2\Im\bigl(\frac{\partial ^2\scrF}{\partial z^i\partial z^j}
     \bigr) dz^i\wedge \overline{dz}^j = \frac {\sqrt{-1}}2\Im(\tau _{ij})
     dz^i\wedge \overline{dz}^j. \tag{1.16} $$

Formulas \thetag{1.14}--\thetag{1.16} are standard in the literature on
special \Kah\ geometry; they show that our global \theprotag{1.1}
{Definition} reproduces the usual local characterization.  We term~$\{z^i\}$
a {\it special coordinate system\/}.  We characterize special coordinate
systems below in \theprotag{1.37} {Definition}.

        \remark{\protag{1.17} {Remark}}
 Condition~\thetag{1.2} does {\it not\/} mean that the complex structure
$I$~is flat.  Indeed, if $\nabla I=0$, then $M$~is a flat \Kah\ manifold,
locally isometric to~$\CC^n$.  Such a manifold is special \Kah, but of a very
special type.  Note that the existence of a flat symplectic structure has
nontrivial global topological consequences but gives no local restriction.
Equation~\thetag{1.2}, on the other hand, is a stringent local condition.
        \endremark

        \remark{\protag{1.18} {Remark}}
 Based on an earlier version of this paper, Zhigin Lu~\cite{L} proved that
there are no nonflat {\it complete\/} special \Kah\ manifolds.
        \endremark

        \remark{\protag{1.19} {Remark}}
 The special \Kah\ condition~\thetag{1.7} automatically implies that
$\nabla$~is torsionfree, since \thetag{1.4}~is twice the real part
of~\thetag{1.7}.  
        \endremark

        \remark{\protag{1.20} {Remark}}
 Locally, we may specify a  \Kah\ geometry by giving a holomorphic
function~$\scrF(z^1,\dots ,z^n)$ such that  
  $$ \Im\bigl(\frac{\partial ^2\scrF}{\partial z^i\partial z^j} \bigr) >0
      $$
is positive definite.  The function 
  $$ \scrF(z^1,\dots ,z^n) = \frac{\sqrt{-1}}{2} \bigl((z^1)^2+\dots +(z^n)^2
     \bigr)  $$
leads to the flat metric on~$\CC^n$.  A nontrivial example in one dimension
is provided by the holomorphic function 
  $$ \scrF(\tau ) = \frac{\tau ^3}{6},  $$
defined on the upper half plane 
  $$ \HH = \{\tau :\Im\tau >0\}.  $$
The corresponding \Kah\ form 
  $$ \omega =\frac{\sqrt{-1}}{2} \Im(\tau )d\tau \wedge \overline{d\tau}
      $$
has Gauss curvature $1/2(\Im\tau )^3$.  Note that the coordinate conjugate
to~$\tau $ is $w=\partial \scrF/\partial \tau = \tau ^2/2$.  An
adapted\footnote{See \theprotag{1.37} {Definition}.} flat Darboux coordinate
system~$\{x,y\}$ is $x=\Re\tau ,\,y=-\Re\tau ^2/2$.  In these coordinates the
Riemannian metric is
  $$ g=\frac{2(x^2+y)\,dx^{2} + 2x\,dx dy +
     dy^{2}}{\sqrt{x^2+2y}}.  $$
It is the Hessian of the function $\phi =\frac 13 (x^2+2y)^{3/2}$; see
\theprotag{1.24} {Proposition} below.  This metric is incomplete; see
\theprotag{1.18} {Remark}. 
        \endremark

        \remark{\protag{1.21} {Remark}}
 Nowhere do we use the positive definiteness of~$\omega $.  Hence our
discussion applies also to {\it pseudo-\Kah\/} manifolds.  (A pseudo-\Kah\
metric~$\omega $ is nondegenerate and~$d\omega =0$, but it is not assumed
positive definite.) 
        \endremark

We have the following easy result.

        \proclaim{\protag{1.22} {Proposition}}
 \hangindent 29pt\rom(a\rom)\ Let $(M,\omega ,\nabla )$ be a special \Kah\
manifold.  The connection~$\nabla $ determines a horizontal distribution~$H$
in the real cotangent bundle~$T^*M$.  Then $H$~is invariant under the complex
structure of~$T^*M$. \par 
 \hangindent 29pt\rom(b\rom)\ The $(0,1)$~part of the connection~$\nabla $ on
the complex tangent bundle~$TM$ equals the $\dbar$~operator.
        \endproclaim

        \demo{Proof}
 (a)\ Choose a flat Darboux coordinate system~$\{x^i,y_j\}$.  Then the local
1-forms $dx^i,dy_j$ define sections of $T^*M\to M$ whose image is an integral
manifold of~$H$.  Since $dx^i$~and $dy_j$~are the real parts of holomorphic
differentials (see~\thetag{1.11}) their graphs are complex submanifolds. 
 
 (b)\ From~\thetag{1.12} we compute that $\nabla \,\partial /\partial z^i$ is
a form of type~$(1,0)$: 
  $$ \nabla \vect{z^j} = -\frac 12 \frac{\partial \tau _{j\ell }}{\partial
     z^k} dz^k\otimes \vect{y_\ell }, \tag{1.23} $$
Since $\partial /\partial z^i$~is a local basis of holomorphic sections, the
desired assertion follows.
        \enddemo

The Riemannian metric has a very special form in flat real coordinates---it
is the Hessian of a function.  This observation is due to Nigel Hitchin. 

        \proclaim{\protag{1.24} {Proposition}}
 Let $(M,\omega ,\nabla )$ be a special \Kah\ manifold.  Suppose $\{u^\alpha
\}$~is a $\nabla $-flat coordinate system.  (For example, it may be a flat
{\it Darboux\/} coordinate system.)  Then the Riemannian metric~$g$ is 
  $$ g=\frac{\partial ^2\phi }{\partial u^\alpha \partial u^\beta
     }\,du^\alpha \otimes du^\beta  $$
for some real function~$\phi $.  In fact, $\phi $~is a \Kah\ potential.
        \endproclaim

        \demo{Proof}
 In these coordinates the symplectic form $\omega =\frac 12 \omega _{\alpha
\beta }\,du^\alpha \wedge du^\beta $ has constant coefficients.  Now 
  $$ g_{\alpha \beta } = \omega _{\alpha \gamma }I^\gamma _\beta  
     $$
and the special \Kah\ condition~\thetag{1.2} implies 
  $$ \frac{\partial g_{\alpha \beta }}{\partial u^\gamma } = \frac{\partial
     g_{\alpha \gamma }}{\partial u^\beta }.  $$
Hence $g_{\alpha \beta }= \partial \phi _\alpha /\partial u^\beta $ for some
function~$\phi _\alpha $.  The symmetry of~$g_{\alpha \beta }$ now implies
that $\phi _\alpha =\partial \phi /\partial u^\alpha $ for some~$\phi $, as
desired.  
 
To see that $\phi $~is a \Kah\ potential, we compute 
  $$ \aligned
      \sqrt{-1}\,\partial \dbar\phi &= -\frac 12 dId\phi \\
      &= -\frac 12 d\bigl(\frac{\partial \phi }{\partial u^\alpha }I^\alpha
     _\gamma du^\gamma \bigr) \\
      &=-\frac 12 \bigl(\frac{\partial ^2\phi }{\partial u^\alpha \partial
     u^\beta }I^\alpha _\gamma +\frac{\partial \phi }{\partial u^\alpha
     }\frac{\partial I^\alpha _\gamma }{\partial u^\beta } \bigr)du^\beta
     \wedge du^\gamma \\
      &= -\frac 12g_{\alpha \beta }I^\alpha _\gamma du^\beta \wedge
     du^\gamma\\
      &=\omega .\endaligned $$
We use the special \Kah\ condition to pass from the third line to the fourth.
        \enddemo

We next express the special \Kah\ condition~\thetag{1.2} in terms of
coordinates.

        \proclaim{\protag{1.25} {Proposition}}
 Let $(M,\omega )$~be a \Kah\ manifold of dimension~$n$ and $\nabla $~ a flat
torsionfree symplectic connection.  Suppose $\{z^i\}$~is any local
holomorphic coordinate system on~$M$ and $\{x^i,y_j\}$ a flat Darboux
coordinate system.  Write
  $$ \vect{z^i} = \frac 12\bigl(\sigma ^j_i\vect{x^j}- \tau _{ij}\vect{y_j}
     \bigr)  $$
for functions~$\sigma ^j_i, \tau _{ij}$.  Then $d_\nabla I=0$ if and only if
$\sigma ^j_i,\tau _{ij}$~are holomorphic functions of~$z^1,\dots ,z^n$ and 
  $$ \frac{\partial \sigma ^j_i}{\partial z^k} = \frac{\partial \sigma
     ^j_k}{\partial z^i}, \qquad \frac{\partial \tau _{ij}}{\partial z^k} =
     \frac{\partial \tau _{kj}}{\partial z^i}.  $$
        \endproclaim

\flushpar
 The proof is straightforward: Compute $d_\nabla (\proj) = d_{\nabla
}\bigl(dz^i\otimes \vect{z^i} \bigr)$.  Notice that $\tau _{ij}$~is not
necessarily symmetric, but rather $\tau\mstrut _{ik}\sigma ^k_j$~is symmetric
in~$i,j$.
 
There is a holomorphic cubic form~$\Xi $ on a  special \Kah\ manifold
which encodes the extent to which $\nabla $~fails to preserve the 
complex structure.  Namely, set 
  $$ \Xi =-\omega \bigl(\proj,\nabla \proj \bigr) \in
     H^0(M,\Sym^3T^*M). \tag{1.26} $$
That $\Xi $~is symmetric follows from the fact that $\omega $~is
skew-symmetric, $\nabla \omega =0$, and the special \Kah\
condition~\thetag{1.7} (which says that $\nabla \proj$~is symmetric).  The
holomorphicity follows from the computation~\thetag{1.28} below.  Note the
alternative local expression
  $$ \Xi =-\omega (\nabla \zeta , \nabla ^2\zeta ),\tag{1.27} $$
where $\zeta $~is a local complex vector field satisfying~\thetag{1.8}.  We
compute~\thetag{1.26} in special coordinates~$\{z^i\}$ introduced above.
From~\thetag{1.23} and the fact that $\omega $~has type~$(1,1)$, we have
  $$ \split
      \Xi &= -\omega \bigl(dz^i\otimes \vect{z^i} \;,\; \nabla (dz^j\otimes
     \vect{z^j}) \bigr) \\
       &= -dz^i\otimes dz^j \;\;\omega \bigl(\frac 12(\vect{x^i} - \tau _{im}
     \vect{y_m}) \;,\; -\frac 12 \frac{\partial \tau _{j\ell }}{\partial z^k}
     \;dz^k\otimes \vect{y_\ell } \bigr) \\
       &= \frac 14 \frac{\partial \tau _{ji}}{\partial z^k} \;dz^i\otimes
     dz^j\otimes dz^k \\
       &= \frac 14 \frac{\partial^3 \scrF}{\partial z^i\partial z^j\partial
     z^k}\;dz^i\otimes dz^j\otimes dz^k .\endsplit \tag{1.28} $$
Here we use~\thetag{1.14} as well.  The cubic form~$\Xi $ can also be
used\footnote{I learned this from the account in~\cite{BCOV}, though it also
appears in many other works.} to relate the  special \Kah\
connection~$\nabla $ go the Levi-Civita connection~$D$.  Write
  $$ \nabla =D+\AR, \tag{1.29} $$
where $\AR\in \Omega ^1(M,\End_{\RR}TM)$.  Then since $D\proj=0$, we have 
  $$ \Xi =-\omega (\proj,[\AR,\proj]). \tag{1.30} $$
Moreover, there is a complex tensor
  $$ A\in \Omega ^{1,0}\bigl(\Hom(TM,\overline{TM}) \bigr) \tag{1.31} $$
with 
  $$ \AR = A + \Abar. $$ 
To see this, note from~\thetag{1.29} and \theprotag{1.22(ii)} {Proposition}
that $A_\xi $~vanishes on vectors of type~$(0,1)$ if $\xi $~is of
type~$(1,0)$.  Then, since $A_\xi $~is infinitesimal symplectic, for $\zeta
$~of type~$(1,0)$ and $\bar{\eta }$~of type~$(0,1)$, we have 
  $$ \omega (A_\xi \zeta ,\bar\eta ) = -\omega (\zeta ,A_\xi \bar\eta ) =0. $$
Since $\omega $~has type~$(1,1)$, this implies that $A_\xi \zeta $~is of
type~$(0,1)$.  Therefore, $A$~is as claimed in~\thetag{1.31}.  Furthermore,
$A$~and $\Xi $~determine each other.  In particular, we recover the special
\Kah\ structure from~$\Xi $.
 
Conversely,  can start with a smooth cubic form $\Xi \in
C^{\infty}(M,\Sym^3T^*M)$ and ask for the conditions on~$\Xi $ which ensure
that $\nabla $~as defined by~\thetag{1.29} and~\thetag{1.30} is a 
special \Kah\ structure.  Note `$T^*M$'~denotes the complex tangent bundle;
we assume~$\Xi $ to be complex multilinear.  The symmetry of~$\Xi $ implies
that $\nabla $~is symplectic, torsionfree, and satisfies~\thetag{1.2}.
Setting the curvature of~$\nabla $ to zero from~\thetag{1.29} yields the
equation 
  $$ 0=R + d_DA + A\wedge \Abar+\overline{A}\wedge A,  $$
where $R$~is the curvature of the \Kah\ metric on~$M$.  Here `$d_D$'~is the
alternation of the Levi-Civita covariant derivative.  Notice that as
endomorphisms of the tangent bundle $R+A\wedge \Abar+\overline{A}\wedge A$ is
complex linear, whereas $d_DA$~is complex antilinear~\thetag{1.31}; whence
these separately vanish.  The (1,1)~piece of~$d_DA$ is~$\dbar A$, from which
it follows that $\Xi $~is holomorphic.  The remaining equations are
  $$ \aligned
      \partial _DA &=0 \\ 
      R &= -(A\wedge \Abar+\overline{A}\wedge A). \endaligned \tag{1.32} $$
In any local coordinate system~$\{z^i\}$ we write 
  $$ \aligned
      \omega &= \sqrt{-1}\, h_{i\bar{j}} \;dz^i\wedge \overline{dz^j}
     \\
      R &= \pmatrix R^j_{ik\bar{\ell }}\endpmatrix_{i,j}\;dz^k\wedge
     \overline{dz^\ell }\\
      \Xi &= \Xi _{ijk}\; dz^i\otimes dz^j\otimes dz^k \\
      (A_i)^{\bar{k}}_j &= \sqrt{-1}\,\Xi _{ij\ell }h^{\ell
     \bar{k}}.\endaligned $$
As usual, set $R_{i\bar j k\bar\ell }=h_{m\bar j}R^m_{ik\bar\ell }$.  Then
\thetag{1.32}~is 
  $$ \aligned
      D_iA_j &= D_jA_i, \\
      R_{i\bar{j}k\bar{\ell }} &= -h^{\alpha \bar{\beta }}\,\Xi
     _{ik\alpha } \,\overline{\Xi _{j\ell {\beta }}}.\endaligned
       \tag{1.33} $$

We summarize this discussion as follows.

        \proclaim{\protag{1.34} {Proposition}}
  \hangindent 29pt\rom(a\rom)\ If $(M,\omega ,\nabla )$~is a  special
\Kah\ manifold, then there is an associated holomorphic cubic form~$\Xi \in
H^0(M,\Sym^3T^*M)$, defined in~\thetag{1.26}, which
satisfies~\thetag{1.32}.\par 
 \hangindent 29pt\rom(b\rom)\ If $(M,\omega )$~is a \Kah\ manifold and $\Xi
\in H^0(M,\Sym^3T^*M)$ holomorphic cubic form which satisfies~\thetag{1.32},
then $\nabla =D+A$ is a  special \Kah\ structure, where $D$~is the
Levi-Civita connection and $A$~is defined from~$\Xi $ by~\thetag{1.30}. 
        \endproclaim

        \remark{\protag{1.35} {Remark}}
 Lu~\cite{L} noticed that as a consequence of~\thetag{1.33} any special \Kah\
manifold~$M$ has nonnegative scalar curvature~$\rho $: 
  $$ \aligned
      \rho  &= -4h^{i\bar j}h^{k\bar\ell }R_{i\bar jk\bar\ell } \\ 
      &= 4h^{i\bar j}h^{k\bar\ell }h^{\alpha \bar\beta }\,\Xi _{ik\alpha
     }\overline{\Xi _{j\ell \beta }} \\ 
      &= 4|\Xi |^2.\endaligned \tag{1.36} $$
Then he computes~$\triangle\rho $ and uses a maximum principle to argue
that if $M$~is complete, then $\rho =0$, from which $\Xi $ and then
$R$~vanish. 
        \endremark

Next, we discuss special coordinates.

        \definition{\protag{1.37} {Definition}}
 Let $(M,\omega ,\nabla )$~be a  special \Kah\ manifold.\par 
 \rom(a\rom)\ A holomorphic coordinate system~$\{z^i\}$ is {\it special\/} if
$\nabla \Re(dz^i)=0$.\par 
 \hangindent 29pt \rom(b\rom)\ We say that special coordinates~$\{z^i\}$ and
flat Darboux coordinates~$\{x^i,y_j\}$ are {\it adapted\/} if
$\Re(z^i)=x^i$.\par
 \hangindent 29pt \rom(c\rom)\ Special coordinate systems~$\{z^i\},\{w_j\}$
are said to be {\it conjugate\/} if there exists a flat Darboux coordinate
system~$\{x^i,y_j\}$ such that $\Re(z^i)=x^i$ and $\Re(w_j)= -y_j$.
        \enddefinition

\flushpar
 Given adapted special coordinates~$\{z^i\}$ and flat Darboux
coordinates~$\{x^i,y_j\}$, conjugate special coordinates~$\{w_j\}$ are
determined up to translation by a purely imaginary constant.  For adapted
coordinate systems we have equations~\thetag{1.9}--\thetag{1.16}, but note
that $\{w_j\}$, $\tau _{ij}$, $\scrF$, and~$K$ are not completely determined
by~$\{z^i\}$ and~$\{x^i,y_j\}$.  The following proposition clarifies the
choices involved.

        \proclaim{\protag{1.38} {Proposition}}
 Let $(M,\omega ,\nabla )$~be a  special \Kah\ manifold.\par 
 \hangindent 29pt \rom(a\rom)\ Given a flat Darboux coordinate
system~$\{x^i,y_j\}$ there exists an adapted special coordinate
system~$\{z^i\}$.  Any two choices~$\{z^i\}, \{\tilde{z}^i\}$ satisfy
$z^i=\tilde{z}^i+c^i$ for some purely imaginary constants~$c^i$.\par
 \hangindent 29pt \rom(b\rom)\ Given a special coordinate system~$\{z^i\}$
there exists an adapted flat Darboux coordinate system~$\{x^i,y_j\}$.  Any
two choices differ by a change of variables
  $$ \pmatrix x\\y  \endpmatrix = \pmatrix 1&0\\A&1  \endpmatrix
     \pmatrix \tilde{x}\\\tilde{y}  \endpmatrix + \pmatrix 0\\b
     \endpmatrix,  $$
where $A$~is a (real) symmetric matrix and $b\in \RR^n$.\par 
 \hangindent 29pt \rom(c\rom)\ Given a special coordinate system~$\{z^i\}$
the holomorphic prepotential~$\scrF$ is determined up to a change
  $$ \scrF \longrightarrow  \scrF + \frac 12A_{ij}z^iz^j + B_iz^i + C,
      $$
where $A=(A_{ij})$~is a real symmetric matrix, and $B_i,C\in \CC$.  So the
conjugate coordinate system~$\{w_j\}$ is determined up to a change 
  $$ w_j \longrightarrow  w_j + A_{jk}z^k + B_j  $$
and the \Kah\ potential~\thetag{1.15} is determined up to a change
  $$ K \longrightarrow K + \Im(B_i\bar{z}^i).  $$
\par 
 \hangindent 29pt \rom(d\rom)\ If $\{z^i\},\{w_j\}$ are conjugate special
coordinate systems, then any other pair $\{\tilde{z}^i\},\{\tilde{w}_j\}$ of
conjugate special coordinate systems are related by
  $$ \pmatrix z\\w \endpmatrix = P\pmatrix \tilde{z}\\\tilde{w} \endpmatrix +
     \pmatrix a\\b \endpmatrix,\qquad P\in Sp(2n;\RR),\quad a,b\in
     \CC^n. \tag{1.39} $$
The corresponding matrices~$\tau ,\tilde{\tau }$ are related by
  $$ \tau =(D\tilde{\tau } + C) (B\tilde{\tau }+A)\inv , \tag{1.40} $$
where $P=\left(\smallmatrix A&B\\C&D  \endsmallmatrix\right)$. 
        \endproclaim

 \newpage
 \head
 \S{2} The Associated Hyperk\"ahler Manifold
 \endhead
 \comment
 lasteqno 2@  5
 \endcomment

In this section we prove the following theorem, which (in local form) is due
to Cecotti, Ferrara, and
Girardello~\cite{CFG}.\footnote{Equation~\thetag{B.7} in~\cite{CFG}
corresponds to our description of the metric in~\thetag{2.4}, where
their~$Z^I$ are special coordinates on~$M$ and $\{Z^I,W_J\}$ the induced
coordinate system on~$T^*M$.  Then \thetag{B.8b}~describes the flat
connection~$\nabla $.}

        \proclaim{\protag{2.1} {Theorem}}
 The cotangent bundle~$T^*M$ of a  special \Kah\ manifold~$(M,\omega
,\nabla )$ carries a canonical \hyp\ structure. 
        \endproclaim

\flushpar
 Recall that a Riemannian manifold~$(Y,g)$ is \hyp\ if it carries a triple of
{\it integrable\/} almost complex structures~$I,J,K$ which satisfy the
quaternion algebra and such that the associated 2-forms 
  $$ \omega _T(\xi_1 ,\xi_2 ) = g(\xi_1 ,T\xi_2 ),\qquad T=I,J,K, \tag{2.2} $$
are {\it closed\/}.  A useful lemma of Hitchin~\cite{H,p.~64} asserts that if
$\omega _I,\omega _J,\omega _K$~are closed, then $I,J,K$~are integrable.  If
we consider $(Y,\omega _I)$~as a \Kah\ manifold with complex structure~$I$,
then
  $$ \eta =\omega _J + i\omega _K \tag{2.3} $$
is a holomorphic symplectic form. 

        \demo{Proof}
 Consider first a hermitian vector space~$V$ with complex structure~$I$.  The
hermitian metric~$\langle \cdot ,\cdot \rangle$ determines a metric and
symplectic form on the underlying real vector space~$V_\RR$:
  $$ \langle \xi_1 ,\xi_2   \rangle = g(\xi_1 ,\xi_2 ) + i\omega (\xi_1 ,\xi_2
     ),\qquad \xi_1 ,\xi_2 \in V_\RR.  $$
Then $W=V\oplus V^* \cong V\oplus \Vbar$ has a constant \hyp\ structure.  The
complex structure~$J$ is the antilinear map
  $$ \aligned
      J\:V\oplus \Vbar &\longrightarrow V\oplus \Vbar \\
      v_1\oplus \overline{v_2}&\longmapsto -v_2\oplus
     \overline{v_1}.\endaligned  $$
Now define $K=IJ$.  Then $I,J,K$~satisfy the quaternion algebra.  The metric
on~$W_\RR$ is
  $$ g_W(\xi_1 \oplus \alpha_1 ,\xi_2 \oplus \alpha_2 ) = g(\xi_1 ,\xi_2 ) +
     g\inv (\alpha_1 ,\alpha_2 ),\qquad \xi_1 ,\xi_2 \in V_\RR,\quad \alpha_1
     ,\alpha_2 \in V_\RR^*. \tag{2.4} $$
The forms $\omega _I,\omega _J,\omega _K$ are now determined
by~\thetag{2.2}.  It is straightforward to check that the holomorphic
symplectic form~$\eta $ defined in~\thetag{2.3} is the canonical form
on~$W=V\oplus V^*\cong T^*V$: 
  $$ \eta (v_1\oplus \ell_1,v_2\oplus \ell_2 ) = \ell_1(v_2) - \ell_2 (v_1),
     \qquad v_1,v_2\in V,\quad \ell_1,\ell_2 \in V^*.  $$
 
Now let $(M,\omega ,\nabla )$ be  special \Kah\ and let $Y=T^*M$.
Consider the distribution of horizontal spaces on~$Y$ given by the
connection~$\nabla $.  Here `horizontal' means relative to the projection map
$\pi \:Y\to M$.  The horizontal space~$H_y$ at~$y\in Y$ is a complex subspace
of~$T_yY$ by \theprotag{1.22} {Proposition}.  The projection~$\pi $ identifies
$H_y\cong T_mM$, where $m=\pi (y)$, and so the splitting into horizontal and
vertical is a splitting 
  $$ T_yY \cong T_mM \oplus  T^*_mM. \tag{2.5} $$
The linear algebra of the preceding paragraph gives global
endomorphisms~$I,J,K$ which satisfy the quaternion algebra.  According to
Hitchin's lemma to check that this determines a \hyp\ structure we must only
verify that $\omega _I,\omega _J,\omega _K$~are closed.  First, since the
canonical holomorphic symplectic form~$\eta $ on~$Y=T^*M$ is closed,
equation~\thetag{2.3} implies that $\omega _J$~and $\omega _K$~are also
closed.  To see that $\omega _I$~is closed we choose a flat Darboux
coordinate system~$\{x^i,y_j\}$ on an open set~$U\subset M$.  This induces a
local coordinate system $\{x^i,y_j;q_i,p^j\}$ on~$\pi \inv U\subset Y$.
Since the splitting~\thetag{2.5} is induced by~$\nabla $, and $dx^i,dy_j$ are
$\nabla $-flat by definition, it follows that
  $$ \omega _I = dx^i\wedge dy_i + dq_i\wedge dp^i.  $$
This form is closed.
        \enddemo

 \newpage
 \head
 \S{3} Integrable Systems
 \endhead
 \comment
 lasteqno 3@  8
 \endcomment

In the mathematical description of a (finite dimensional) classical
mechanical system one meets a symplectic manifold~$X$ and a Hamiltonian
function.  It is an integrable system if there is a maximal set of Poisson
commuting conserved momenta which includes the Hamiltonian.  Under suitable
hypotheses this leads to a foliation of~$X$ by lagrangian
tori~\cite{GS,\S44}.  The complex analogue leads to the following
definition~\cite{DM1}, which we explain in the succeeding paragraphs.

        \definition{\protag{3.1} {Definition}}
 An {\it algebraic integrable system\/} is a holomorphic map $\pi \:X\to M$
where \par 
 \rom(a\rom)\ $X$~is a complex symplectic manifold with holomorphic
symplectic form~$\eta \in \Omega ^{2,0}(X)$;\par 
 \rom(b\rom)\ The fibers of~$\pi $ are compact lagrangian submanifolds, hence
affine tori;\par 
 \hangindent 29pt\rom(c\rom)\ There is a family of smoothly varying
cohomology classes~$[\rho_m] \in H ^{1,1}(X_m)\cap H^2(X_m;\ZZ),\;\mathbreak
m\in M$, such that $[\rho _m]$~is a positive polarization of the fiber~$X_m$.
Hence $X_m$~is an abelian torsor.\par
        \enddefinition

\flushpar
  The hypothesis that the fibers are compact lagrangian leads to the
conclusion that they are affine tori.  The fact that they are abelian torsors
is an extra hypothesis.  We assume that $X$~and $M$~are smooth.\footnote{The
singularities contain crucial physics, but for the geometry in this section
we restrict to smooth points.}

We now explain this definition and some consequences.  Recall that a
single\footnote{For convenience we use the same notation for the single
abelian varieties in this explanatory paragraph as we do in the rest of the
text for families of abelian varieties.}  {\it abelian variety\/} is a
quotient~$A=V/\Lambda $ of a complex vector space~$V$ by a full real
lattice~$\Lambda $ such that $H^{1,1}(A)\cap H^2(A;\ZZ)\not= \nomathbreak 0$
and there is a positive class~$[\rho ]$ in this intersection.  Such a class
is called a {\it polarization\/} and is represented by a unique invariant
positive closed $(1,1)$-form~$\rho $ on~$A$.  The polarization is {\it
principal\/} if $\int_{A}\frac{\rho ^n}{n!}=1$.  Note that $\rho $~is a real
symplectic form on~$A$, and since it is invariant it is a symplectic form
on~$V_\RR$ as well.  Also, since $\rho $~is an integral class, it induces a
symplectic form on~$\Lambda \cong H_1(A)$.  Let $\{\gamma ^i,\delta
_j\}\subset \Lambda $ be a symplectic basis.  Then there is a unique basis
$\{\omega _i\}$ of holomorphic differentials on~$A$ with
  $$ \int_{\gamma _i}\omega _j = \delta ^i_j, \tag{3.2} $$
where $\delta ^i_j$~is the Kronecker symbol.  In fact, we can identify
$\{\omega _i\}$~as the complex basis of~$V^*$ dual to~$\{\gamma ^i\}$.  Now
  $$ \int_{\delta _j}\omega _i = \tau _{ij} \tag{3.3} $$
defines the {\it period matrix\/}~$\tau $ of~$A$.  The Riemann bilinear
relations state that the matrix $\tau =(\tau _{ij})$ belongs to the Siegel
upper half space 
  $$ \HH_n = \{\text{$\tau $~an $n\times n$ complex matrix}: \text{$\tau$ is
     symmetric and $\Im\tau$ is positive definite}\}.  $$
The group~$Sp(2n;\RR)$ acts transitively on~$\HH_n$.  A change of symplectic
basis $\{\gamma ^i,\delta _j\}$ transforms~$\tau $ by an element of a
discrete subgroup~$\Gamma \subset Sp(2n;\RR)$ which depends on the
polarization. (For a principal polarization~$\Gamma =Sp(2n;\ZZ)$.)  An
abelian torsor~$X$ is a principal homogeneous space for an abelian
variety~$A=V/\Lambda$ with a polarization~$[\rho ]$.  Here $V$~is the space
of invariant vector fields on~$X$ and $\Lambda \subset V$~the lattice of such
vector fields which exponentiate to the identity map.  We can identify~$A$ as
the Albanese variety of~$X$.  Any point~$x\in X$ determines an isomorphism
$A\to X$, and the pullback of~$[\rho ]$ is a polarization~$[\hat{\rho }]$
of~$A$ which is independent of the choice of~$x$.  The period matrix of~$X$
is equal to the period matrix of~$A$.
 
An algebraic integrable system $\pi \:X\to M$ leads to a parametrized version
of the preceding discussion.  First, the holomorphic symplectic form~$\eta $
gives an isomorphism
  $$ i\:T^*M \overset\cong \to\longrightarrow  V,  $$
where $V\to M$ is the bundle of invariant vector fields along the fibers
of~$\pi $.  For a complex function $f\:M\to \CC$ and complex vector
field~$\xi $ on~$X$ we have $\eta \bigl(i(df),\xi \bigr) = \pi ^*df(\xi )$.
This leads to a fiberwise action of~$T^*M\cong V$ by exponentiation.  Let
$\Lambda $~be the kernel of the action.  A basic fact is that $\Lambda $~is a
complex lagrangian submanifold of~$T^*M$, where $T^*M$~has the canonical
holomorphic symplectic structure.  (See~\cite{GS,\S44} for proofs of the
assertions made here.)  Furthermore, $\Lambda $~intersects each fiber
of~$\cotM$ in a full lattice.  The quotient~$A=\cotM/\Lambda $ is a family of
abelian varieties parametrized by~$M$; it is the bundle of Albanese varieties
of $X\to M$.  Since $\Lambda $~is complex lagrangian, the canonical
holomorphic symplectic form on~$T^*M$ passes to a holomorphic symplectic
form~$\etah$ on the quotient~$A$.  Now a local lagrangian section of $\pi
\:X\res U\to U$ over an open set~$U\subset M$ induces a local isomorphism
$X\res U\cong A\res U$, and this isomorphism maps~$\etah$ to~$\eta $.  Such
sections may not exist globally.  Since any two choices of local section lead
to isomorphisms which differ by a translation on each fiber, the family of
polarizations~$[\rho _m]$ on~$X\to M$ define a family of
polarizations~$[\rhoh_m]$ on~$A\to M$.
 
To summarize: Every algebraic integrable system $X\to M$ has a canonically
associated algebraic integrable system $A \to M$ whose fibers are abelian
varieties.  (An analogous assertion holds for real integrable systems.)
Either system determines a well-defined {\it period map\/}
  $$ \tau \:M\longrightarrow \scrA_n = \HH_n/\Gamma   $$
into the moduli space~$\scrA_n$ of suitably polarized abelian varieties.
 
Now the bundle of lattices~$\Lambda $ determines a flat connection~$\nabla $
on~$\cotM$, hence also on~$TM$.  Since $\Lambda $~is lagrangian, $\nabla $~
is torsionfree.  Also, the polarization~$[\rhoh_m]$ on~$A_m=T^*_mM/\Lambda
_m$ determines a real symplectic form on~$T^*_mM$ which restricts to an
integral symplectic form on the lattice~$\Lambda _m$.  The dual
2-form~$\omega $ on~$M$ is flat---$\nabla \omega =0$---and since $\nabla $~is
torsionfree it follows that $\omega $~is closed.  Thus $\omega $~is a real
symplectic form on~$M$.  The holonomy group of the flat connection~$\nabla $
is contained in the {\it integral\/} symplectic group~$Sp(\Lambda ^*_m)$ at
each~$m\in M$, where $\Lambda ^*_m$~is the dual lattice to~$\Lambda _m$.
Furthermore, by the definition of a polarization $\omega $~is a (positive
definite) \Kah\ form on~$M$.  If $\{\gamma ^i,\delta _j\}$ is a local
symplectic basis of sections of~$\Lambda \subset T^*M$, then we can write
  $$ \omega =\gamma ^i\wedge \delta _i.  $$
There is also a global formula for~$\omega $.  First, each
polarization~$[\rhoh_m]$ is represented by a unique invariant closed form
$\rhoh_m\in \Omega ^{1,1}(A_m)$.  The family of forms~$\{\rhoh_m\}$ is flat
with respect to~$\nabla $.  Now the connection~$\nabla $ on~$\cotM$ induces
an integrable distribution of horizontal planes on~$A$, and we
extend~$\{\rhoh_m\}$ to a form~$\rhoh\in \Omega ^{1,1}(A)$ by requiring that
$\rhoh$~vanish on those horizontal planes.  Then $d\rhoh=0$.  The global
formula for~$\omega $ is expressed in terms of~$\rhoh$ and the holomorphic
symplectic form $\etah\in \Omega ^{2,0}(A)$: 
  $$ \omega =\frac{1}{4}\int_{A/M} \etah \wedge \bar{\etah }\wedge
     \frac{\rhoh ^{n-1}}{(n-1)!}.  $$

The conclusion of this discussion is a result stated by Donagi and
Witten~\cite{DW}.

        \proclaim{\protag{3.4} {Theorem}}
 \par\hangindent29pt \rom(a\rom)\ Let $(X\to M,\eta ,[\rho_m] )$ be an
algebraic integrable system.  Then the \Kah\ form~$\omega $ and the
connection~$\nabla $ constructed above comprise a  special \Kah\
structure on~$M$.  Furthermore, there is a lattice $\Lambda ^*\subset TM$
whose dual $\Lambda \subset T^*M$ is a complex lagrangian submanifold, and
the holonomy of~$\nabla $ is contained in the integral symplectic group
defined by~$\Lambda ^*$.\par
 \hangindent29pt \rom(b\rom)\ Conversely, suppose $(M,\omega ,\nabla )$~is a
 special \Kah\ manifold.  Suppose further that there is a lattice
$\Lambda ^*\subset TM$, flat with respect to~$\nabla $, whose dual $\Lambda
\subset \cotM$ is a complex lagrangian submanifold.  Then $A=\cotM/\Lambda
\to M$ admits a canonical holomorphic symplectic form~$\eta $ and a family of
polarizations~$[\rho _m]$ so that $(A\to M,\eta ,[\rho _m])$ is an algebraic
integrable system whose fibers are abelian varieties.
        \endproclaim

\flushpar

        \remark{\protag{3.5} {Remark}}
  The lattice~$\Lambda $ in~(b) may be specified by a covering of
distinguished flat Darboux coordinate systems~$\{x^i,y_j\}$ whose transition
functions satisfy~\thetag{1.6} with~$P\in Sp(2n;\ZZ)$.  In this case we also
restrict the allowable special coordinate systems~$\{z^i\}$ by requiring that
$\{\Re(z^i)\}$~be part of a distinguished flat Darboux coordinate system. 
        \endremark

        \demo{Proof}
 For part~(a) it remains to verify the  special \Kah\
condition~\thetag{1.2}, or equivalently~\thetag{1.7}.  We work locally.  Let
$\{\gamma ^i,\delta _j\}$~be a local symplectic basis of sections of~$\Lambda
$.  Since $\gamma ^i,\delta _j$~are closed 1-forms we can find  flat Darboux
coordinates~$\{x^i,y_j\}$ so that $\gamma ^i=dx^i$ and $\delta _j=dy_j$.  Now
$\gamma ^i,\delta _j$~also determine families of cycles on~$A$ and we can
find holomorphic functions~$z^i,w_j$ such that 
  $$ dz^i=\int_{\gamma ^i}\etah,\qquad dw_j=-\int_{\delta _j}\etah. 
     $$
Here the integrals are over the families of cycles in the fibration $A\to M$,
and Stokes' theorem shows that the integrals are holomorphic $(1,0)$-forms.
It is easy to check that $\Re(dz^i)=dx^i$ and $\Re(dw_j)=-dy_j$, so we can
arrange that $\Re(z^i)=x^i$ and $\Re(w_j)=-y_j$.  Then 
  $$ \zeta =\frac 12 \bigl(z^i\vect{x^i} - w_j\vect{y_j} \bigr)  $$
is a local complex vector field which satisfies~\thetag{1.8}.  This
implies~\thetag{1.7}.
 
Notice that the vector fields $\omega _i=\vect{z^i}$ define local holomorphic
differentials on the fibers of~$A\to M$, and they satisfy~\thetag{3.2}.  Thus
equation~\thetag{3.3} defines the period matrix~$(\tau _{ij})$ relative
to~$\{\gamma ^i,\delta _j\}$.  Equations~\thetag{3.2} and~\thetag{3.3} are
equivalent to equation~\thetag{1.12}: 
  $$ \vect{z^i} = \frac 12\bigl(\vect{x^i}- \tau _{ij}\vect{y_j}
     \bigr).  $$
 
By now the proof of~(b) should be clear.  Given $(M,\omega ,\nabla ,\Lambda
)$, the family of polarizations on $A=\cotM/\Lambda \to M$ is represented by
the dual of the \Kah\ form~$\omega $.  Hence $A\to M$ is a family of abelian
varieties.  The symplectic form is induced from the canonical symplectic form
on~$\cotM$.  The hypothesis that $\Lambda $~is complex lagrangian makes the
quotient~$\cotM/\Lambda $ complex symplectic. 
        \enddemo

        \remark{\protag{3.6} {Remark}}
 An arbitrary family of abelian varieties $A\to M$ does {\it not\/} admit a
symplectic form.  For that the differential of the period map must come from
a cubic form $c\in H^0(M,\Sym^3\cotM)$.  (See~\cite{DM1,\S7}.)  Here we assume
a given identification of the bundle~$V$ with~$\cotM$.  (Recall that $V$~is
the bundle of constant vector fields along the fibers of $A\to M$.)  The
cubic condition on the period matrix is essentially the  special \Kah\
condition~\thetag{1.2}, as is clear from \theprotag{1.25} {Proposition}.  Of
course, the cubic form is~\thetag{1.26}.
        \endremark

        \remark{\protag{3.7} {Remark}}
 The preceding discussion applies to the pseudo-\Kah\ case with one
modification:  the polarization classes~$[\rho _m]$ are no longer positive
definite.  So $X_m$~is an affine torus with an indefinite polarization.  We
term this an {\it indefinite algebraic integrable system.\/}
        \endremark

The discussion in~\S{2} applies directly to the quotient~$T^*M/\Lambda $, and
so \theprotag{2.1} {Theorem} yields the following. 

        \proclaim{\protag{3.8} {Theorem}}
 Let $(X\to M, \eta ,[\rho _m])$ be an algebraic integrable system.  Then
$X$~carries a canonical \hyp\ structure. 
        \endproclaim

 \newpage
 \head
 \S{4} Projective Special \Kah\ Manifolds
 \endhead
 \comment
 lasteqno 4@  9
 \endcomment

We term the triple $(M,L,\omega )$ a {\it Hodge manifold\/} if $(M,\omega
)$~is \Kah\ and $L\to M$ is a holomorphic hermitian line bundle with
curvature\footnote{Since we do not use so many indices in this section, we
revert to the standard notation $i=\sqrt{-1}$.} $-\tpi\omega $.  This implies
$[\omega ]\in H^2(M;\RR)$ is an integral class.  We begin with a geometric
lemma about the principal \Cx bundle $\pi \:\Mtil\to M$ obtained by deleting
the zero section from $L\to M$.  First, the hermitian connection on~$L$ is
also a connection on $\pi \:\Mtil\to M$, that is, a \Cx-invariant
distribution of horizontal subspaces.  Also, the bundle $\pi ^*L\to\Mtil$ has
a canonical nonzero holomorphic section~$s$. 

        \proclaim{\protag{4.1} {Lemma}}
 Let $\omtil\in \Omega ^{1,1}(\Mtil)$ denote the form which equals~$|s|^2\pi
^*\omega $ on pairs of horizontal vectors, vanishes on a horizontal vector
paired with a vertical vector, and is $-1/\pi $~times the canonical \Kah\
form on pairs of vertical vectors.  Then 
  $$ \omtil=\frac{i}{2\pi }\dbar\partial |s|^2. \tag{4.2} $$
Thus $d\omtil=0$, which implies that $\omtil$~is a pseudo-\Kah\ metric
on~$\Mtil$ of Lorentz type.  Finally,
  $$ \pi ^*\omega =\frac{i}{2\pi }\dbar\partial \log |s|^2. \tag{4.3} $$
        \endproclaim

\flushpar
 The canonical \Kah\ form on a hermitian line~$L$ is $\frac i2\partial
\dbar|s|^2,\; s\in L$.  The metric~$\omtil$ is negative definite on
fibers and positive definite on horizontal subspaces.  It has
signature~$(n,1)$ where $n=\dim M$. 

        \demo{Proof}
 Let $t$ be a nonzero holomorphic section of $L\res U\to U$ for an open
set~$U\in M$, and set $h(z)=|t(z)|^2,\;z\in U$.  We use local coordinates
$\langle z,\lambda \rangle\mapsto \lambda\, t(z)\in \pi \inv U\subset \Mtil$,
where~$\lambda \in \CC^\times $.  Now $s(z,\lambda )= \lambda t(z)$, and so
$|s(z,\lambda )|^2 = |\lambda |^2h(z)$.  Compute the right hand side
of~\thetag{4.2}.  To verify the description of~$\omtil$ given
before~\thetag{4.2}, note that $\xi -\lambda h\inv \partial h(\xi
)\vect{\lambda }$ is the horizontal lift of a tangent vector~$\xi $ in~$U$.
Formula~\thetag{4.3} is the standard curvature formula for the hermitian
connection.
        \enddemo

The usual definition for what we call a projective special \Kah\ structure is
a particular type of {\it variation of Hodge structure\/}, which was
considered specifically in a paper of Bryant and Griffiths~\cite{BG}.  We
discuss this first and defer our description to \theprotag{4.6(b)}
{Proposition}.  Our version of the usual definition emphasizes the fact that
the parameter space is a Hodge manifold, but it is equivalent to the
definition in~\cite{BG} (cf.,~\cite{C} for the relationship to~\cite{St}).

        \definition{\protag{4.4} {Definition}}
  \par\hangindent29pt \rom(\rom i\rom)\ A {\it projective special \Kah\
structure\/} on an $n$~dimensional Hodge manifold $(M,L,\omega )$ is a \break
triple $(V,\nabla ,Q)$ where\par
 \hangindent52pt\hskip24pt \rom(a\rom)\ $V\to M$ is a holomorphic vector bundle of rank~
$n+1$ with a given holomorphic inclusion $L\hookrightarrow V$;\par 
 \hangindent52pt\hskip24pt \rom(b\rom)\ $\nabla $~is a flat connection on the
underlying real bundle $V_\RR\to M$ such that $\nabla (L)\subset V$ and the
section 
  $$ \aligned
      M &\longrightarrow \PP\bigl[(V_\RR)_{\CC}\bigr] \\ 
      m &\longmapsto L_m\endaligned \tag{4.5} $$
is an immersion with respect to~$\nabla $;\par 
 \hangindent52pt\hskip24pt \rom(c\rom)\ $Q$~is a nondegenerate skew form
on~$V_\RR$ which has type~(1,1) with respect to the complex structure and
satisfies~$\nabla Q=0$.  Furthermore, we assume that $Q\res{L\times
\overline{L}}$ is $i/2\pi $ times the hermitian metric on~$L$.\par
 \hangindent29pt \rom(\rom i\rom i\rom)\ An {\it integral projective special \Kah\
structure\/} is a quadruple $(\Lambda ,V,\nabla ,Q)$ with $(V,\nabla ,Q)$ as
in~\rom(\rom i\rom) and $\Lambda \subset V_{\RR}$ a flat submanifold which
intersects each fiber in a full lattice such that $Q\res{\Lambda \times
\Lambda }$~has integral values.
        \enddefinition

\flushpar
 In this definition $\nabla $~and $Q$~are extended to the
complexification~$(V_\RR)_\CC$ of~$V_\RR$.  The flat connection gives a local
identification of~$V_\RR$---hence also of its complexification~$(V_\RR)_\CC$
and the projectivization~$\PP\bigl[(V_\RR)_\CC\bigr]$---with any fiber.  The
immersion condition in~(b) states that $m\mapsto L_m$ is an immersion into
the $2n+1$~dimensional projective space of a local trivialization
of~$\PP\bigl[(V_\RR)_\CC\bigr]$. 
 
The data in~(ii) define a variation of polarized Hodge structures of weight~3
with Hodge numbers~$h^{3,0}=1,\; h^{2,1}=n$ with an extra immersion
condition.  This is the form of the definition in~\cite{BG}.
(See~\cite{CGGH} for the basic definitions related to variations of Hodge
structures.)  We recover the Hodge filtration~$\{F^p\}$ by setting $F^3=L$,
$F^2=V$, $F^1={F^3}^\perp$, and $F^0=(V_\RR)_\CC$.  (Here `${}^\perp$' is
with respect to~$Q$.)  The Griffiths transversality condition $\nabla
(F^3)\subset F^2$ is given in~(b) above; the condition $\nabla (F^2)\subset
F^1$ follows from this and the immersion condition~\cite{BG,pp.~82--83}.
\theprotag{4.6} {Proposition} below implies that $iQ\res{H^{2,1}\times
\overline{H^{2,1}}}$ is positive definite, where $H^{2,1}= F^2\cap
\overline{F^1 }$.  Variations of Hodge structure without the lattice, as
in~(i), were considered in~\cite{S}.
 
Our main observation in this section is the following.  We prefer to take the
structure in~(b) as the definition of projective special \Kah.

        \proclaim{\protag{4.6} {Proposition}}
 Let $(M,L,\omega )$~be a Hodge manifold with associated pseudo-\Kah\
manifold~$(\Mtil,\omtil)$ and canonical section~$s$.\par
 \hangindent 29pt\rom(a\rom)\ A projective special \Kah\ structure on
~$(M,L,\omega )$ induces a \Cx-invariant special pseudo-\Kah\
structure~$\nabtil$ on~$(\Mtil,\omtil)$ with $\nabtil s=\proj$.\par
 \hangindent 29pt\rom(b\rom)\ Conversely, a \Cx-invariant special
pseudo-\Kah\ structure~$\nabtil$ on~$(\Mtil,\omtil)$ which satisfies $\nabtil
s=\proj$ induces a projective special \Kah\ structure on~$(M,L,\omega )$.
        \endproclaim

\flushpar
 Recall that $\omtil$~is defined in \theprotag{4.1} {Lemma}.   The
canonical section~$s$ defined there can be viewed as the holomorphic vertical
vector field on~$\Mtil$ induced by the \Cx~action.

        \demo{Proof}
 (a)\ Let $\nabtil=\pi ^*\nabla $ be the lifted flat connection on~$\pi ^*V$.
Using the inclusion $L\hookrightarrow V$ we view $s$~as a section of~$\pi
^*V$.  The immersion condition~\thetag{4.5} implies that
  $$ \nabtil s\:T\Mtil \longrightarrow \pi ^*V \tag{4.7} $$
is an isomorphism.  (Note that $\nabtil s\subset \pi ^*V$ by the Griffiths
transversality in~(b).)  Using the real isomorphism underlying~\thetag{4.7}
we obtain a real flat connection on~$\Mtil$; we also denote it by~`$\nabtil$'.
Furthermore, under~\thetag{4.7} the form $\Qtil = \pi ^*Q$ pulls back
to~$-\omtil$.  This follows by differentiating the equation
  $$ \frac{i}{2\pi } |s|^2 = \Qtil(s,\sbar),  $$
assumed in~(c), to obtain 
  $$ \omtil = \frac{i}{2\pi }\dbar\partial |s|^2 = -\Qtil(\nabtil
     s,\overline{\nabtil s}).  $$
Thus $\nabtil\omtil=0$.  Now under~\thetag{4.7} the section~$s$ corresponds
to a holomorphic vector field~$\zeta $ which satisfies~\thetag{1.8}.  This
proves that $\nabtil$~satisfies the  special \Kah\
condition~\thetag{1.7}, and by \theprotag{1.19} {Remark} $\nabtil$~is also
torsionfree. 
 
(b)\ We simply indicate the construction of~$(V,\nabla ,Q)$.  First, let
$V$~be the quotient of~$T\Mtil$ by the \Cx~action.  Then $V$~is a holomorphic
bundle over~$M$, and the inclusion of vertical vectors in~$T\Mtil$ induces an
inclusion $L\hookrightarrow V$.  The connection~$\nabtil$ on~$(T\Mtil)_{\RR}$
induces a connection~$\nabla $ on~$V_{\RR}$; the immersion condition in
\theprotag{4.4(i)(b)} {Definition} follows from the hypothesis $\nabtil s =
\proj$.  The form~$-\omtil$ on~$\Mtil$ induces a skew form~$Q$ on~$V$.
        \enddemo

\flushpar
 Notice as a consequence of~(c) and the description of~$\omtil$ in
\theprotag{4.1} {Lemma} that $iQ\res{H^{2,1}\times \overline{H^{2,1}}}$ is
positive definite.  

Now the discussion of special coordinates, holomorphic prepotential,
etc\. from~\S{1} applies to $(\Mtil,\omtil,\nabtil)$.  We make
\Cx-equivariant choices on~$\Mtil$ and consider the induced tensors on~$M$.
We work on~$\pi \inv (U)$ for $U\subset M$ a sufficiently small open set.  We
do not choose Darboux coordinates, but only a flat local symplectic
framing\footnote{It is denoted $\{\vect{x^i},\vect{y_j}\}$ in~\S{1}, but here
we do not consider coordinate functions~$x^i$ and~$y_j$.} of~$T\Mtil$, which
we require to be \Cx-invariant.  We say that a complex tensor field
on~$\Mtil$ has {\it degree~$n$\/} if it transforms under~$\lambda
\in\CC^{\times }$ by multiplication by~$\lambda ^n$.  The vector field~$\zeta
$ (which corresponds to~$s$ under~\thetag{4.7}) has degree~1.  So
from~\thetag{1.9} we see that a special coordinate function~$z^i$ also has
degree~1.  In other words, $z^i$~is a local holomorphic section of~$L\to M$.
Thus a special coordinate system~$\{z^i\}$ on~$\Mtil$ gives rise to local
{\it projective\/} coordinates on~$M$ (which transform as sections of~$L$).
From~\thetag{1.13} we see that the period matrix~$(\tau _{ij})$ is a scalar,
and from~\thetag{1.14} that the holomorphic prepotential~$\scrF$ has
degree~2, i.e., $\scrF$~is a local holomorphic section of~$L^{\otimes 2}$.
Because of the \Cx-invariance there is less flexibility in choosing~$\{z^i\}$
and~$\scrF$ than in the nonprojective case---different choices differ by a
{\it homogeneous\/} function.

        \proclaim{\protag{4.8} {Proposition}}
 Let $(M,L,\omega ,\nabtil)$ be a projective special \Kah\ manifold.\par
 \hangindent 29pt \rom(a\rom)\ Given a special projective coordinate
system~$\{z^i\}$ the holomorphic prepotential~$\scrF$ is determined up to a
change
  $$ \scrF \longrightarrow  \scrF + \frac 12A_{ij}z^iz^j,  $$
where $A=(A_{ij})$~is a real symmetric matrix.  Hence the conjugate special
projective coordinate system~$\{w_j\}$ is determined up to a change
  $$ w_j  \longrightarrow w_j + A_{jk}z^k.  $$
\par 
 \hangindent 29pt \rom(b\rom)\ If $\{z^i\},\{w_j\}$ are conjugate special
projective coordinate systems, then any other pair
$\{\tilde{z}^i\},\{\tilde{w}_j\}$ of conjugate special projective coordinate
systems are related by
  $$ \pmatrix z\\w \endpmatrix = P\pmatrix \tilde{z}\\\tilde{w} \endpmatrix,
     \qquad P\in Sp(2n;\RR).  $$
        \endproclaim

From~\thetag{4.3} we see that the lift of the metric~$\omega $ to~$\Mtil$ has
a global ``\Kah\ potential''~$\Ktil$, which we write in special coordinates
as 
  $$ \aligned
      \Ktil &\doteq \mopl |s|^2 \\
       &\doteq \mopl \Qtil(s,\sbar) \\
       &\doteq \mopl \bigl(-\omtil(\zeta ,\bar{\zeta}) \bigr) \\
       &\doteq \mopl \Im(z^i\bar{w}_i) \\
       &\doteq \mopl \Im\bigl(z^i\frac{\partial \overline{\scrF}}{\partial
     \overline{z^i}}\bigr) .\endaligned  $$
Here `$\doteq$'~means `equals up to an additive constant'.  $\Ktil$~pulls
down to a local \Kah\ potential on~$M$ via a local holomorphic section of
$\pi \:\Mtil\to M$. 
 
The cubic form~$\tilde\Xi $ of the  special \Kah\ structure on~$\Mtil$
(see~\thetag{1.26}) is a holomorphic section
  $$ \Xi \in H^0(M,Sym^3T^*M\otimes L^{\otimes 2}),  $$
as follows easily from~\thetag{1.28}.  It is a basic ingredient in the
analysis of~\cite{BG}, where it is derived from an {\it infinitesimal\/}
variation of Hodge structure.  Since $\zeta $~is holomorphic of type~$(1,0)$,
we have $\omega (\zeta ,\nabtil\zeta )=0$, and by differentiating $\omega
(\zeta ,\nabtil^2\zeta )=0$.  Differentiating once more we conclude
from~\thetag{1.27} that the cubic form in this case is
  $$ \Xi =\omega (\zeta ,\nabtil^3\zeta ) = \Qtil(\nabla ^3s,s).  $$
 
We can use $\tilde{\Xi }\in H^0(\Mtil,\Sym^3T^*\Mtil)$ and the associated
$\Atil\in \Omega ^{1,0}\bigl(\Hom(T\Mtil,\overline{T\Mtil}) \bigr)$ to
introduce an algebra structure on $T\Mtil\otimes _\RR\CC$.  Fix $\tilde{m}\in
\Mtil$ and denote $V=T_{\tilde{m}}\Mtil$.  It is easy to see that
$\Atil$~vanishes on~$\zeta $, and it is a well-defined map $W\otimes
W\to\overline{W}$, where $W$~is the orthogonal complement to~$\zeta $.
(Under the projection~$\pi $ we can identify $W\cong T_mM$, where $m=\pi
(\tilde{m})$.)  We now obtain a graded algebra~$C$: Set $C_0=\CC\cdot \zeta
$, $C_1=W$, $C_2=\overline{W}$, and $C_3=\CC\cdot \bar{\zeta}$; then $\zeta
$~acts as the identity, the multiplication $C_1\otimes C_1\to C_2$ is given
by~$\Atil$, and the multiplication $C_1\otimes C_2\to C_3$ is $\alpha \otimes
\bar{\beta }\mapsto \omega (\alpha ,\bar{\beta })\bar{\zeta }$.
Associativity is trivial to verify.
 
Now we consider the implications of the lattice~$\Lambda \subset V_\RR$ in an
{\it integral\/} projective special \Kah\ structure on~$M$.  Under the
isomorphism~\thetag{4.7} the lift $\pi ^*\Lambda \subset \pi ^*V_\RR$ induces
a lattice $\Ltil\subset T\Mtil$.  Now $\Ltil$~is $\nabtil$-flat by
hypothesis, so by \theprotag{1.22} {Proposition} it is a complex submanifold.
Locally $\Ltil$~is the graph of a $\nabtil$-flat vector field on~$\Mtil$, so
the dual~$\Ltil^*$ is locally the graph of a $\nabtil$-flat 1-form.  Since
$\nabtil$~is torsionfree, this 1-form is also holomorphic and so
$\Ltil^*\subset T^*\Mtil$ is complex lagrangian.  Thus \theprotag{3.4(b)}
{Theorem} and \theprotag{3.8} {Theorem} apply to give the following
conclusion.

        \proclaim{\protag{4.9} {Proposition}}
 Suppose $(M,L,\omega ,\nabtil,\Lambda )$~is an integral projective special
\Kah\ manifold of dimension~$n$.  Then there is an associated indefinite
algebraic integrable system $X\to\Mtil$, where $\Mtil$~is $L$~with the zero
section removed.  The total space~$X$ carries a ``pseudo-hyperk\"ahler''
structure of real signature~$(4n,4)$.
        \endproclaim

\flushpar
 The fibers of this integrable system are the intermediate Jacobians
associated to the underlying variation of Hodge structure.  The symplectic
form on this family of intermediate Jacobians was constructed by Donagi and
Markman~\cite{DM2} (for the case of a family of Calabi-Yau manifolds).  The
pseudo-hyperk\"ahler structure was also given by Cort\'es~\cite{C}.  As in
the nonprojective case we restrict our local Darboux framings to lie in the
lattice, and so the matrices~$A,P$ in \theprotag{4.8} {Proposition} must be
integral.

 \newpage
 \head
 \S{5} Remarks on $N=2$~Gauge Theories in Four Dimensions
 \endhead
 \comment
 lasteqno 5@  2
 \endcomment

We make some brief remarks on the role of special \Kah\ manifolds in global
supersymmetric theories.  We do not comment on their role in supergravity.
References for the quantum physics are~\cite{SW1} and~\cite{SW2}.  For a
mathematical development of the relevant classical supersymmetry,
see~\cite{DF}.  The quantum aspects of our discussion have no pretension to
rigor.
 
We first recall the origin of the local formula~\thetag{1.15} for the \Kah\
potential.  It arises from the lagrangian for the complex scalars in the four
dimensional $N=2$~{\it vector multiplets\/}.  There is a superspace
description in terms of the superspace~$N^{4|8}$, which is an extension of
ordinary four dimensional Minkowski space with eight odd dimensions.  The
complexification of the odd distribution splits into two pieces, and there is
a corresponding notion of a {\it chiral\/} map $\Sigma \:N^{4|8}\to\CC$.
Such a map describes an (abelian) $N=2$~vector multiplet.  (More precisely,
it is a component of the curvature of a constrained connection on
superspace.)  The most general supersymmetric lagrangian for $n$ such
multiplets is specified by a holomorphic function $\scrF\:\CC^n\to\CC$.  The
theory is free if $\scrF$~is quadratic.  Upon reduction to $N=1$
superspace~$N^{4|4}$ each multiplet~$\Sigma $ decomposes into an $N=1$~chiral
multiplet~$\Phi $ and an $N=1$~vector multiplet $\scrA$.  The lagrangian for
the chiral multiplets is determined from the \Kah\ potential~$K$, and a
computation gives the formula~\thetag{1.15} for~$K$ in terms of~$\scrF$.
 
Next, we emphasize that a special \Kah\ manifold does {\it not\/} define a
{\it classical\/} field theory for $N=2$~vector multiplets.  We do obtain a
classical lagrangian from a special coordinate system, as explained in the
previous paragraph.  Furthermore, any \Kah\ manifold~$M$ does determine a
well-defined $N=1$~supersymmetric field theory for a chiral field $\Phi
\:N^{4|4}\to M$.  However, the change of special coordinates~\thetag{1.39}
must be accompanied by a duality\footnote{That is, electromagnetic duality.}
transformation on the gauge field in the vector multiplet~$\scrA$, and this
only makes sense in the {\it quantum\/} theory.  Moreover, this duality
transformation only makes sense when the holonomy of~$\nabla $ is contained
in the {\it integral\/} symplectic group.  Thus a special \Kah\ manifold~$M$
with a lattice as in \theprotag{3.4} {Theorem} determines\footnote{Since
typically $M$~is incomplete this is not yet a full description of a theory.
Also, an abelian gauge theory, which has a positive $\beta $-function, only
makes sense as an effective field theory, not as a fundamental theory.} a
{\it quantum\/} field theory which locally has a semiclassical description in
terms of $N=2$~vector multiplets.  The manifold~$M$ is the moduli space of
quantum vacua.  According to \theprotag{3.4} {Theorem} such a theory is
always specified by an algebraic integrable system.
 
These abelian theories describe the low energy behavior of the Coulomb branch
of nonabelian $N=2$ supersymmetric gauge theories, with or without matter.
The simplest example~\cite{SW1} has gauge group~$SU(2)$ and no matter.  Then
$M$~is the universal curve~$M(2)$ for the modular group~$\Gamma (2)\subset
SL(2;\ZZ)$, which we can identify as~$\CP^1$ with 3~points omitted, say
$M(2)=\CP^1 - \{-1,1,\infty \}$.  The universal curve $X(2)\to M(2)$ is the
algebraic integrable system which defines the model.  Many more examples have
been found, all of course involving integrable systems.  (See~\cite{D} for a
review.)
 
So far we have taken $M$~to be smooth.  As stated above, a nonflat $M$~is not
complete and an honest physical theory is formulated on some
completion of~$M$.  For example, for the pure $SU(2)$~gauge theory the
special \Kah\ metric on the moduli space $\CP^1-\{-1,1,\infty \}$ is complete
near~$\infty $, but the singular points~$-1,1$ are at finite distance.  At
these points other fields are massless and must be added to the low energy
description. 

We now remark further on the physical origin of the lattice~$\Lambda $.  It
is a feature of four dimensional abelian gauge theories; supersymmetry is
irrelevant.  (See~\cite{AgZ,\S3} for a recent discussion.) Consider a four
dimensional gauge theory with gauge group~$G=\TT^n$, where $\TT\cong U(1)$ is
the circle group.  The theory is specified by a complex bilinear form~$\tau $
on the Lie algebra~$\frak{g} $ whose imaginary part~$\Im \tau $ is an inner
product.\footnote{For $n=1$ in the standard basis the form~$\tau $ is
usually written $\tau =\frac{\theta }{\pi } + \frac{8\pi \sqrt{-1}}{e^2}$,
where $e$~is the coupling constant.}  The lagrangian density in Minkowski
space is 
  $$ L = \Bigl\{ -\frac{1}{8\pi }\Im\tau (F_A,*F_A) + \frac{1}{8\pi }\Re\tau
     (F_A,F_A)\Bigr\}\,|d^4x|, \tag{5.1} $$
where $A$~is a connection and $|d^4x|$~the standard density.  There is a
lattice~$\frak{g}_\ZZ\subset \frak{g}$ whose elements exponentiate to the
identity in~$G$, and each basis of this lattice produces a matrix $(\tau
_{ij})\in \HH_n$ which represents the form~$\tau $.  The group~$GL(n;\ZZ)$
permutes these bases.  The larger {\it duality group\/}~$Sp(2n;\ZZ)$ is
generated by this group together with the electromagnetic duality
transformation.  The latter expresses the theory in terms of a ``dual''
connection~$\Atil$ and the bilinear form~$-\tau \inv $.  The lagrangian has
the same form as~\thetag{5.1}, and the operator~$F_A$ in the original theory
corresponds to~$*F_{\Atil}$ in the dual theory.  The action of~$Sp(2n;\ZZ)$
which is generated acts on~$\tau $ by~\thetag{1.40}. 
 
Fix a basis of~$\frak{g}_{\ZZ}$ and so write the curvature as
$F_A=(F_A^i)_{i=1,\dots ,n}$.  There are $n$~electric charges~$q^i$ and
$n$~magnetic charges~$g_i$ for charged matter we might put into the theory.
Classically, the electric charge in a spatial region bounded by a
surface~$\Sigma $ is defined to be
  $$ q^i = \int_{\Sigma } \frac{\sqrt{-1}}{2\pi }*F_A^i,  $$
and the enclosed magnetic charge is 
  $$ g^i = (n_m)^i = \int_{\Sigma } \frac{\sqrt{-1}}{2\pi }F_A^i.  $$
The electric and magnetic charges of a {\it quantum\/} state are computed
from the corresponding operators in the quantum theory.  Now $(n_m)^i$ is an
integer by Chern-Weil theory for the {\it compact\/} gauge group~$\TT^n$.  In
the classical theory $(n_m)^i$~is an integer-valued function on the space of
classical solutions; in the quantum theory it assigns an integer to each
quantum state.  There are other integers~$(n_e)_i$ attached to quantum states
from the Noether charges associated to global infinitesimal gauge
transformations.  Here the integrality is from the fact that certain
exponentials of these infinitesimal transformations are the identity
operator.  These integers are related to the electric charge of states via
the formula
  $$ q^i = \bigl((\Im \tau) \inv \bigr)^{ij}\bigl((\Re\tau )_{jk}(n_m)^k +
     (n_e)_j \bigr). $$
It is convenient to consider the complex quantity
  $$ q^i + \sqrt{-1}g^i = \bigl((\Im \tau) \inv \bigr)^{ij}\bigl(\tau_{jk}
     (n_m)^k + (n_e)_j \bigr). $$
As the $n_m,n_e$ run over all integers, this runs over the points of the {\it
(electromagnetic) charge lattice\/}~$\Lambda ^*$ in~$\CC^n$.  There is an
integral symplectic form~$\omega $ on~$\Lambda ^*$ defined by
  $$ \aligned
      \omega \Bigl( \bigl(\smallmatrix g\\q
     \endsmallmatrix\bigr),\left(\smallmatrix \tilde{g} \\ \tilde{q}
     \endsmallmatrix\right) \Bigr) &= g^i(\Im\frac\tau2 )_{ij}\tilde{q}^j -
     \tilde{g}^i(\Im\frac\tau2 )_{ij}q^j \\
      &= (n_m)^i(\tilde{n}_e)_i - (\tilde{n}_m)^i(n_e)_i.\endaligned
     \tag{5.2} $$
It is preserved by the duality group.  Equation~\thetag{5.2} is the form of
charge quantization usually referred to as the ``Dirac-Schwinger-Zwanziger
condition''. 

Returning to an $N=2$ supersymmetric abelian gauge theory, we have the moduli
space~$M$ of the complex scalars which carries the special geometry we have
been discussing.  There is a distinguished set of conjugate special
coordinate systems related by integral coordinate transformations.  In each
such coordinate system we have a lagrangian description as a gauge theory
with gauge group~$G=\TT^n$ (with distinguished basis for the Lie algebra).
We should regard the~$\CC^n$ where the coordinates live as the complexified
Lie algebra with its distinguished basis.  The electromagnetic charge
lattice~$\Lambda ^*$ discussed in the previous paragraph defines a global
lattice in the complex conjugate cotangent bundle~$\overline{T^*M}\cong TM$.
Note in the notation of~\S1 that $\left(\smallmatrix (n_m)^i\\(n_e)_j
\endsmallmatrix\right)$ transforms analogously to~$\left(\smallmatrix
dx^i\\dy_j \endsmallmatrix\right)$, and $q^i + \sqrt{-1}g^i$ transforms
analogously to~$\overline{dz^i}$.  (See formulas~\thetag{1.6}
and~\thetag{1.12}.)

There is a further geometric input from the {\it BPS mass formula\/}.
Namely, in the classical theory the central charge~$Z$ in the
$N=2$~supersymmetry algebra is a complex-valued locally constant function on
the space of solutions to the classical field equations.  In the quantum
theory (at a point~$m\in M$) it is an operator whose eigenvalues are complex
numbers.  Let $\{z^i\}, \{w_j\}$ be distinguished conjugate special
coordinate systems.  This means that there is a lagrangian description for
the $N=2$~theory in terms of a prepotential~$\scrF(z^1,\dots ,z^n)$ with $w_j
= \partial \scrF/\partial z^j$.  The BPS~formula involves possible additional
$\TT$~charges $S^\alpha $ which may appear in the theory.  These have integer
eigenvalues.  The BPS formula asserts  that the eigenvalue of~$Z$ is
  $$ z^i(n_e)_i + w_i (n_m)^i + s^\alpha \frac{m_\alpha }{\sqrt2} ,
      $$
where $s^\alpha $~is the eigenvalue of~$S^\alpha $ and $(n_e)_i,(n_m)^i$ are
the integers defined above.  Let $\Gamma \subset \CC$ denote the set of
points so described.  If there are no~$S^\alpha $, then the fact that $\Gamma
$~is intrinsic and the transformation law for $\left(\smallmatrix
(n_m)^i\\(n_e)_j \endsmallmatrix\right)$ implies that there is no translation
in the coordinate change~\thetag{1.39} between different sets of
distinguished conjugate coordinate systems.  However, in the presence of
charges~$S^\alpha $ there may be a nonzero translational component.


\newpage


\Refs\tenpoint

\ref
\key AgZ     
\by L. \'Alvarez-Gaum\'e, F. Zamora 
\paper Duality in quantum field theory (and string theory) 
\paperinfo {\tt hep-th/9709180}
\endref

\ref
\key BCOV    
\by M. Bershadsky, S. Cecotti, H. Ooguri, C. Vafa
\paper Kodaira-Spencer theory of gravity and exact results for quantum string
amplitudes 
\jour  Commun. Math. Phys.  
\vol 165  
\yr 1994 
\pages 311--428
\finalinfo {\tt hep-th/9309140}
\endref

\ref
\key BG      
\by R. L. Bryant, P. A. Griffiths 
\paper Some observations on the infinitesimal period relations
for regular threefolds with trivial canonical bundle
\inbook Arithmetic and Geometry, Vol. II
\pages 77--102
\bookinfo Progr. Math., 36 
\publ Birkh\"auser 
\publaddr Boston 
\yr 1983
\endref

\ref
\key CO      
\by P. Candelas, X. C. de la Ossa
\paper Moduli space of Calabi-Yau manifolds 
\jour Nuclear Phys. B  
\vol 355  
\yr 1991 
\pages 455--481
\endref

\ref
\key CGGH    
\by J. Carlson, M. Green, P. Griffiths, J. Harris 
\paper Infinitesimal variations of Hodge structure. I 
\jour Compositio Math. 
\vol 50 
\yr 1983 
\pages 109--205
\endref

\ref
\key CFG     
\by S. Cecotti, S. Ferrara, L. Girardello 
\paper Geometry of type II superstrings and the moduli of superconformal
field theories 
\jour Int. J. Mod. Phys. A 
\vol 4 
\yr 1989 
\pages 2475--2529
\endref

\ref
\key C       
\by V. Cort\'es 
\paper On hyper-K\"ahler manifolds associated to Lagrangian K\"ahler
submanifolds of $T\sp *\CC^n$  
\jour Trans. Amer. Math. Soc.
\vol 350 
\yr 1998 
\pages 3193--3205
\endref

\ref
\key CRTP    
\by B. Craps, F. Roose, W. Troost, A. Van Proeyen
\paper What is special K\"ahler geometry?
\jour Nucl. Phys. B  
\vol 503 
\pages 565--613 
\yr 1997 
\finalinfo {\tt hep-th/9703082}
\endref

\ref 
\key DF 
\by P. Deligne, D. S. Freed 
\paper Supersolutions 
\inbook Quantum Fields and Strings:  A Course for Mathematicians
\vol 1
\publ American Mathematical Society 
\toappear 
\endref

\ref
\key D       
\finalinfo {\tt alg-geom/9705010}
\by R. Y. Donagi
\paper Seiberg-Witten integrable systems
\inbook Algebraic geometry---Santa Cruz 1995 
\pages 3--43 
\vol 62 
\procinfo Proc. Sympos. Pure Math. 
\publ Amer. Math. Soc. 
\publaddr Providence, RI 
\yr 1997
\endref

\ref
\key DM1      
\by R. Y. Donagi, E. Markman 
\paper  Spectral covers, algebraically completely integrable, Hamiltonian
systems, and moduli of bundles 
\inbook Integrable systems and quantum groups (Montecatini Terme, 1993) 
\pages 1--119
\bookinfo Lecture Notes in Math., 1620 
\publ Springer 
\publaddr Berlin 
\yr 1996
\endref

\ref
\key DM2     
\by R. Y. Donagi, E. Markman 
\paper Cubics, integrable systems, and Calabi-Yau threefolds
\pages 199--221
\jour Israel Math. Conf. Proc. 
\vol 9 
\yr 1996 
\paperinfo Proceedings of the Hirzebruch 65 Conference on Algebraic Geometry
(Ramat Gan, 1993)
\endref 

\ref
\key DW      
\by R. Y. Donagi, E. Witten 
\paper Supersymmetric Yang-Mills theory and integrable systems
\jour Nucl. Phys. B 
\vol 460 
\yr 1996 
\pages 299--334
\endref

\ref 
\key F 
\by P. Fr\'e 
\paper Lectures on special K\"ahler geometry and electric-magnetic
duality rotations
\jour Nucl. Phys. Proc. Suppl. 
\vol 45BC 
\yr 1996 
\pages 59--114
\finalinfo {\tt hep-th/9512043}
\endref

\ref 
\key G 
\by S. J. Gates 
\paper Superspace formulation of new non-linear sigma models
\jour J. Nucl. Phys. B 
\vol 238 
\yr 1984 
\page 349--366 
\endref 


\ref
\key GS      
\by V. Guillemin, S. Sternberg 
\book Symplectic techniques in physics
\publ Cambridge University Press
\publaddr Cambridge 
\yr 1990
\endref

\ref
\key H       
\by N. Hitchin 
\book Monopoles, Minimal Surfaces and Algebraic Curves 
\publ Les Presses de L'Universit\'e de Montr\'eal 
\bookinfo S\'eminaire de Math\'ematiques Sup\'erieures 
\vol 105 
\yr 1987
\endref

\ref 
\key L
\by Z. Lu 
\paper A note on the special K\"ahler manifolds 
\paperinfo preprint 
\endref 

\ref 
\key ST 
\by G. Sierra, P. K. Townsend 
\paper An introduction to $N = 2$ rigid supersymmetry 
\inbook Supersymmetry and Supergravity 1983
\ed B. Milewski  
\publ World Scientific  
\publaddr Singapore  
\yr 1983 
\page 396
\endref

\ref
\key SW1     
\by N. Seiberg, E. Witten 
\paper Electric-magnetic duality, monopole condensation, and confinement
in $N=2$ supersymmetric Yang-Mills theory 
\jour Nucl. Phys. B 
\vol 430 
\yr 1994 
\pages 485--486 
\moreref 
\paper Erratum 
\jour Nucl. Phys. B 
\vol 430 
\pages 485--486 
\yr 1994
\finalinfo {\tt hep-th/9407087}
\endref

\ref
\key SW2     
\by N. Seiberg, E. Witten 
\paper Monopoles, duality and chiral symmetry breaking in $N=2$
supersymmetric QCD 
\jour Nucl. Phys. B 
\vol 431 
\yr 1994 
\pages 484--550
\finalinfo {\tt hep-th/9408099}
\endref

\ref
\key SW3     
\by N. Seiberg, E. Witten 
\paper Gauge dynamics and compactification to three dimensions 
\inbook The Mathematical Beauty of Physics: a Memorial Volume for Claude
Itzykson 
\eds J. M. Drouffe, J. B. Zuber
\publ World Scientific
\publaddr Singapore
\yr 1997 
\pages 333--366
\finalinfo {\tt hep-th/9607163}
\endref

\ref
\key S       
\by C. T. Simpson 
\paper Higgs bundles and local systems 
\jour Inst. Hautes \'Etudes Sci. Publ. Math.
\vol 75 
\yr 1992 
\pages 5--95
\endref

\ref
\key St      
\by A. Strominger 
\paper Special geometry 
\jour Commun. Math. Phys. 
\vol 133 
\yr 1990 
\pages 163--180
\endref

\ref
\key WP      
\by B. de Wit, A. Van Proeyen 
\paper  Potentials and symmetries of general gauged $N=2$
supergravity-Yang-Mills models 
\jour Nucl. Phys. B  
\vol 245 
\yr 1984 
\pages 89--117
\endref

\endRefs
\enddocument